\documentclass[journal]{IEEEtran}

\usepackage[pdftex]{graphicx}
\graphicspath{{./figures/}}

\usepackage{comment}
\usepackage[caption=false]{subfig}

\usepackage{hyperref}
\hypersetup{pdfpagemode=UseNone}

\usepackage{multirow} % nehalem for merging cells in a table
\usepackage{array}
\usepackage{cite} % for grouping references
\usepackage{enumitem} % to remove indent from lists (Nehalem)
\usepackage{amsmath} % for formulas (Nehalem)
\usepackage{longtable} % for splitting long table across pages (Nehalem)
\usepackage{url} % for breaking long url in references (Nehalem)

\begin{document}
\title{Robust Smartphone App Identification Via\\Encrypted Network Traffic Analysis}

\author{Vincent~F.~Taylor,~\IEEEmembership{}
        Riccardo~Spolaor,~\IEEEmembership{}
        Mauro~Conti~\IEEEmembership{}
		and Ivan~Martinovic~\IEEEmembership{}
		\thanks{V. F. Taylor and I. Martinovic are with the Department of Computer Science, University of Oxford, Oxford, United Kingdom. E-mail: \mbox{vincent.taylor@cs.ox.ac.uk} ; \mbox{ivan.martinovic@cs.ox.ac.uk}.}
		\thanks{R. Spolaor and M. Conti are with the Dipartimento di Matematica, Università di Padova, Padua 35122, Italy. E-mail: \mbox{conti@math.unipd.it} ; \mbox{rspolaor@math.unipd.it}.}}% <-this % stops a space

%\markboth{IEEE TRANSACTIONS ON INFORMATION FORENSICS AND SECURITY, VOL. 11, NO. 1, JANUARY 2016}%



\maketitle

\begin{abstract}
The apps installed on a smartphone can reveal much information about a user, such as their medical conditions, sexual orientation, or religious beliefs. Additionally, the presence or absence of particular apps on a smartphone can inform an adversary who is intent on attacking the device. In this paper, we show that a passive eavesdropper can feasibly identify smartphone apps by fingerprinting the network traffic that they send. Although SSL/TLS hides the payload of packets, side-channel data such as packet size and direction is still leaked from encrypted connections. We use machine learning techniques to identify smartphone apps from this side-channel data. In addition to merely fingerprinting and identifying smartphone apps, we investigate how app fingerprints change over time, across devices and across different versions of apps. Additionally, we introduce strategies that enable our app classification system to identify and mitigate the effect of ambiguous traffic, i.e., traffic in common among apps such as advertisement traffic. We fully implemented a framework to fingerprint apps and ran a thorough set of experiments to assess its performance. We fingerprinted 110 of the most popular apps in the Google Play Store and were able to identify them six months later with up to 96\% accuracy. Additionally, we show that app fingerprints persist to varying extents across devices and app versions.
\end{abstract}

%\begin{IEEEkeywords}
%Cellular phones, information security, privacy.
%\end{IEEEkeywords}

\IEEEpeerreviewmaketitle

\section{Introduction}

Smartphone usage continues to grow explosively, with Gartner reporting consumer purchases of smartphones as exceeding one billion units in 2014, up~28.4\% over 2013~\cite{gartnersmartphonesales}. Mobile analytics company, Flurry, reports that app usage in 2014 grew by 76\%~\cite{flurrystats}. Nielson reports that for Q4 2014, U.S. smartphone users accessed 26.7 apps per month, spending more than~37 hours using them~\cite{nielsonmanyapps}. The Guardian reports that smartphones are now the most popular way to access the internet in the UK~\cite{guardianalexhern}. Additionally, The Telegraph reports that smartphone-generated mobile traffic is roughly twice as much as PCs, tablets, and mobile routers combined~\cite{telegrapheightfold}. This combination of increased app usage and significant amounts of app traffic places the smartphone in the spotlight for anyone looking to understand the usage of specific apps by the general public.

Smartphone users typically install and use apps that are in line with their interests. Apps cover a broad spectrum of functionality such as medical, finance, entertainment, and lifestyle. As a result, the apps installed on typical smartphones may reveal sensitive information about a user's medical conditions, hobbies, and sexual/religious preferences~\cite{seneviratne2014predicting}. An adversary could also infer who a user banks with, what airline they usually fly on, and which company provides them insurance. This information may be particularly useful in ``spear phishing attacks". In addition to uncovering the aforementioned high-level information, an adversary can also use app identification to enumerate and exploit potentially vulnerable apps in an attempt to gain privileges on a smartphone.

Network traffic fingerprinting is not a new area of research, and indeed the literature exemplifies techniques for network traffic classification on traditional computers~\cite{nguyen2008survey}. On smartphones, however, app fingerprinting and identification is frustrated in several ways. Port-based fingerprinting fails because apps deliver their data predominantly using HTTP/HTTPS. Typical web page fingerprinting fails since apps usually send data back and forth using text formats such as XML and JSON, thus removing rich information (such as the number of files and file sizes) that aid web page classification. Additionally, many apps use content delivery networks (CDNs) and third-party services, thus eliminating hostname resolution or IP address lookup as a viable strategy. Observing (DNS) address resolution or TLS handshakes also proves less useful due to the use of CDNs. Moreover, DNS and TLS exchanges may not be observed at all due to the use of client-side caching, or simply due to the mobile nature (i.e., transient connectivity) of smartphones.

In this paper, we focus on understanding the extent to which smartphone apps can be fingerprinted and later identified by analysing the encrypted network traffic coming from them. We exploit the fact that while SSL/TLS protects the payload of a packet, it fails to hide other coarse information revealed by network traffic patterns, such as packet lengths and direction. Additionally, we evaluate the robustness of our app fingerprinting framework by measuring how it is affected by different devices, different app versions, or the mere passage of time. In what follows, we motivate the utility of app fingerprinting and identification by outlining four concrete scenarios where it may be useful.

% verbatim text STARTS here (not so verbatim anymore april 2017)
\textbf{Attackers targeting specific apps.} An adversary in possession of exploits (perhaps zero-day exploits) for particular apps may use app fingerprinting to identify these vulnerable apps on a network. The adversary can build a fingerprint of a vulnerable app (or vulnerable version of an app) ``offline" and then later use it to identify these apps in the wild. Once a vulnerable app has been identified, the adversary may then exploit whatever vulnerabilities it contains for their own benefit. It is particularly worrying to consider an adversary fingerprinting vulnerable mobile banking apps as a precursor to launching an attack. By performing app fingerprinting, the adversary increases their accuracy when targeting victims, and becomes more discreet when attacking, by having the ability to launch their attack only against vulnerable devices.

\textbf{Attackers targeting specific users.} Within the wireless network a victim is connected to, an adversary could surreptitiously monitor the victim's network traffic to identify what apps they were using or had installed on their device. The disclosure of such private information by an adversary may considerably harm some high-profile victims. For example, a competing political candidate may gain an advantage by revealing to the public that a married opponent was using a dating/flirting app on his/her device. The gravity of this problem is highlighted when one considers the advanced persistent threat (APT) context where high-profile persons are specifically targeted. 

\textbf{Network management.} App fingerprinting provides valuable data about the types of apps and usage patterns of these apps within an organisation. In the current era of bring-your-own-device (BYOD), this information would be invaluable to network administrators wanting to optimize their networks. For example, knowing the most popular apps and their throughput and latency requirements for good user experience, administrators could then configure quality of service on their network such that particular apps performed more efficiently. Additionally, app fingerprinting may be used to determine whether disallowed apps were being used on an enterprise network. The administrator could then take appropriate action against the offender.

\textbf{Advertising and market research.} Companies can rely on app fingerprinting techniques as a source of information to aid market research. Suppose an analytics company wants to know the popularity of apps in a particular location or during a particular event (e.g., during a music concert). This company could potentially fingerprint apps and then go into their location of interest to identify app usage from within a crowd of users. By fingerprinting app usage within their target population, advertisers may be able to build better profiles of their target market and consequently be better able to deploy targeted advertising.\\
%verbatim text ENDS here (not so verbatim anymore april 2017)

\noindent{In this paper we extend AppScanner, first presented by the authors in~\cite{7467370}, along several important dimensions. AppScanner is a highly-scalable and extensible framework for the fingerprinting and identification of apps from their network traffic. The framework is encryption-agnostic, and only analyses side-channel data, thus making it perform the same whether network traffic is encrypted. AppScanner was tested with Android apps and devices, but due to the similarity in app network communication across platforms, we believe that it can be easily ported to work with other mobile operating systems. We make the following contributions to the state-of-the-art beyond the original paper:\\}

\begin{enumerate}
\item Implementation of a novel machine learning strategy that can be used to identify ambiguous network traffic that is similar between apps. Ambiguous network traffic includes advertisement traffic, third-party library traffic, and other common web-API traffic. Such traffic would hinder classification performance in the system described in our earlier paper, because training data would sometimes have conflicting labels. With the improved system, ambiguous traffic can be identified and handled accordingly.

\item An analysis of the robustness of app fingerprinting across different devices and app versions. We also analyse the time invariability of app fingerprints, my measuring how performance is affected when attempting to identify apps using fingerprints generated six months earlier.

\item Evidence that app fingerprints (are in many cases) time, app version, and device invariant. This lends support to the idea of being able to use app classification in real-world settings, since it suggests that fingerprints persist to varying extents.\\
\end{enumerate}

\noindent{The rest of the paper is organised as follows: Section II surveys related work and positions our contribution within the literature; Section III overviews how our system works at a high-level and explains key terminology; Section IV outlines our approach to identifying ambiguous network flows that reduce system performance; Section V overviews the various datasets that were collected; Section VI evaluates performance under a variety of scenarios; Section VII discusses ways of improving classifier accuracy using post-processing strategies; Section VIII discusses our observations throughout this work; and finally Section IX concludes the paper.}
\section{Related Work}
\label{related-work}

Much work has been done on analysing traffic from workstations and web browsers~\cite{hintz2003fingerprinting}. At first glance, fingerprinting smartphone apps may seem to be a simple translation of existing work. While there are some similarities, such as end-to-end communication using IP addresses/ports, there are nuances in the type of traffic sent by smartphones and the way in which it is sent that makes traffic analysis in the realm of smartphones distinct from traffic analysis on traditional workstations~\cite{6356057, 6676399, 6077033, FeghhiL16tifs}. With this in mind, we outline related work by first enumerating traffic analysis approaches on workstations (Section~\ref{ta-workstation}), and then focusing on traffic analysis on smartphones (Section~\ref{ta-smartphone}).

\subsection{Traditional Traffic Analysis on Workstations}
\label{ta-workstation}

Traditional analysis approaches have relied on artefacts of the HTTP protocol to make fingerprinting easier. For example, when requesting a web page, a browser will usually fetch the HTML document and all corresponding resources identified by the HTML code such as images, JavaScript and style-sheets. This simplifies the task of fingerprinting a web page since the attacker has a corpus of information (IP addresses, sizes of files, number of files) about the various resources attached to an individual document.

Many apps, for scalability, build their APIs on top of content delivery networks (CDNs) such as Akamai or Amazon AWS~\cite{forbescdn}. This reduces (on average) the number of endpoints that apps communicate with. In the past, it may have been useful to look at the destination IP address of some traffic and infer the app that was sending the traffic. Presently, requests to \textit{graph.facebook.com}, for example, may possibly be from the Facebook app, but they may also be from a wide range of apps that query the Facebook Graph API. With the advent of CDNs and standard web service APIs, more and more apps are sending their traffic to similar endpoints and this frustrates attempts to fingerprint app traffic based on destination IP address only.

In the literature, several works considered strong adversaries (e.g., governments) that may leverage traffic analysis. Those adversaries are able to capture the network traffic flowing through communication links~\cite{Raymond:2001:TAP:371931.371972}. Liberatore et al.~\cite{Liberatore:2006:ISE:1180405.1180437} showed the effectiveness of proposals aiming to identify web-pages via encrypted HTTP traffic analysis. Subsequently, Herman et al.~\cite{Herrmann:2009:WFA:1655008.1655013} outperformed Liberatore et al. by presenting a method that relies on common text mining techniques to the normalized frequency distribution of observable IP packet sizes. This method correctly classified some~97\% of HTTP requests. Similar work was proposed by Panchenko et al.~\cite{Panchenko:2011:WFO:2046556.2046570}. Their proposal correctly identified web pages despite the use of onion routing anonymisation such as Tor. More recently, Cai et al.~\cite{cai2012touching} presented a web page fingerprinting attack and showed its effectiveness despite traffic analysis countermeasures (e.g., HTTPOS).

Unfortunately, the aforementioned work was not designed for smartphone traffic analysis. Indeed, the authors focused on identifying web pages on traditional desktop computers and leverage the fact that the HTTP traffic can be very unique depending on the structure of the web page. Despite smartphone apps communicating using HTTP as well, they usually rely on text-based APIs, the usage of which removes rich traffic features that would otherwise be present in typical HTTP traffic. For this reason, fingerprinting network traffic on smartphones is a more complicated process.

\subsection{Traffic Analysis on Smartphones}
\label{ta-smartphone}

In early work on the topic, Dai et al.~\cite{Dai2013} propose NetworkProfiler, an automated approach to profiling and identifying Android apps using dynamic methods. They use user-interface fuzzing (UI fuzzing) to automatically explore different activities and functions within an app, while capturing and logging the resulting network traffic. The authors inspect HTTP payloads in their analysis and thus this technique only works with unencrypted traffic. Given the overall trend towards encrypting network communications, this approach will become less useful over time. Dai et al. did not have the full ground truth of the traffic traces they were analysing, so it is difficult to systematically quantify how accurate NetworkProfiler was in terms of precision, recall, and overall accuracy.

St\"{o}ber et al.~\cite{Stober2013} propose a scheme for identifying entire devices using characteristic traffic patterns coming from the devices. They contend that~70\% of smartphone traffic belongs to background activities happening on the device and that this can be leveraged to create a fingerprint. The authors posit that 3G transmissions can be realistically intercepted and demodulated to obtain side channel information such as the amount of data and timing information. The authors leverage `bursts' of data from which to generate their identification since they cannot analyse the TCP payload directly. Using supervised learning algorithms, the authors build a model of the background traffic coming from devices. This model is then capable of identifying data from similar background traffic at a later time. The authors conclude that using approximately~15 minutes of captured traffic can result in a classification accuracy of over~90\%. A major drawback with this work is that the system needs six hours of training and 15~minutes of monitoring to achieve reliable fingerprint matching.

Wang et al.~\cite{wangwificns} propose a system for identifying smartphone apps from encrypted~802.11 frames. They collect data from target apps by running them dynamically and training classifiers with features from the Layer~2 frames that were observed. This work shows promise, but suffers from the fact that the authors only test 13 arbitrarily chosen apps from eight distinct app store categories and collect network traces for only five minutes. Indeed, the authors discover that longer training times have an adverse effect on accuracy when classifying some apps with their system. Moreover, the authors use an insufficient sample size (i.e., only 13 apps) to validate their results. By taking into account a large set of apps in our earlier work~\cite{7467370}, we show how increasing the number of apps negatively influences classifier accuracy. It is problematic to quantify Wang et al.'s results, in general, since they have no way to collect accurate ground truth, i.e., a labelled dataset that is free of noise from other apps. Indeed, our methodology minimises noise by running a single app at a time, and we still had to filter~13\% of the traffic collected because it was background traffic from other apps. AppScanner solves the aforementioned problems by using a larger sample of apps from a wider set of categories and collecting network traffic for substantially more time.

Conti et al.~\cite{conti2014can} and Saltaformaggio et al.~\cite{198409} identify specific actions that users are performing within their smartphone apps. Due to similarity, we briefly describe the approach of Conti et al. The authors identify specific actions through flow classification and supervised machine learning. Their system works in the presence of encrypted connections since the authors only leverage coarse flow information such as packet direction and size. The authors achieved more than~95\% accuracy for most of the considered actions. This work suffers from its specificity in identifying discrete actions. By choosing specific actions within a limited group apps, Conti et al. may benefit from the more distinctive flows that are generated. Their system also does not scale well since a manual approach was taken when choosing and fingerprinting actions. Indeed, the authors chose a small set of apps and a subset of actions within those apps to analyse.

Our prior work~\cite{7467370} improves on the weaknesses of the systems described above. First, by leveraging only side-channel information, we are able to classify apps in the face of encrypted network traffic. Additionally, our system is trained and tested on 110 apps with traffic collected from each app for 30~minutes. Due to the nature of our framework, apps can also be trained automatically, removing the need for human intervention.

Our prior work is however limited in its handling of ambiguous traffic. Ambiguous traffic, i.e., traffic that is common among more than one apps, would frustrate our previous system and cause poorer performance. Our prior work also does not provide an understanding of the variability and longevity of app fingerprints. In this work, we measure how different devices, app versions, or the passage of time affects app fingerprinting.
\section{System Overview}
\label{system-design}

As an overview, AppScanner fingerprints smartphone apps by using machine learning to understand the network traffic that has been generated by them. Patterns in app generated traffic, when later seen, are used to identify the app.

Unfortunately, apps sometimes have traffic patterns in common because they share libraries, such as ad libraries, that generate similar traffic\footnote{Traffic generated by libraries will typically be common among apps using that particular library.} across distinct apps. This can frustrate attempts at app classification using traffic analysis, since it may generate false positives. Thus, a strategy is needed to first identify traffic that is shared among apps, so that it can be appropriately labelled before being passed to classifiers. We call traffic shared among apps \textit{ambiguous traffic} and the remaining traffic \textit{distinctive traffic}.

\begin{figure}[]
\centering
\includegraphics[width=3.5in, clip=true, trim=0.35in 0.3in 4.4in 2.3in]{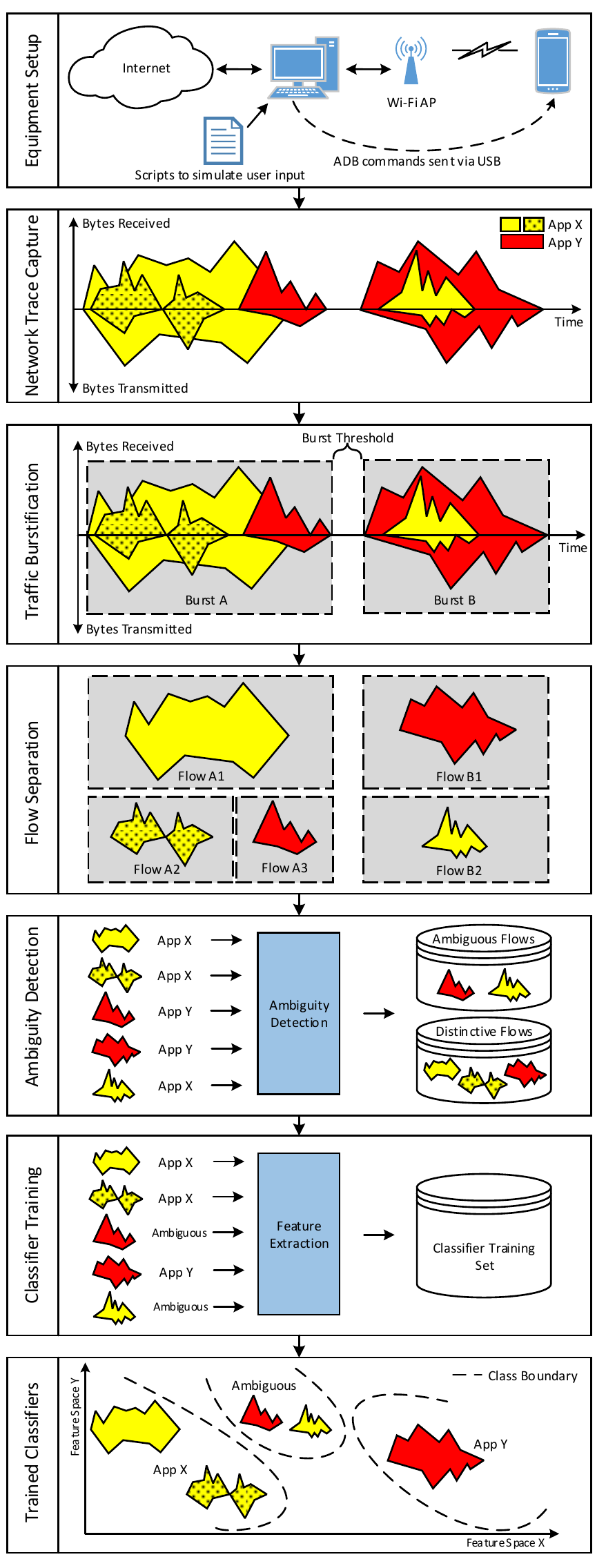}
\caption{High-level representation of classifier training, and a visualisation of bursts and flows within network traffic.}
\label{fig_combined_outline}
\end{figure}

Central to our fingerprinting methodology is the concept of a \textit{burst} and a \textit{flow}. We define these important terms below:\\

\textbf{Burst:} A burst is the group of all network packets (irrespective of source or destination address) occurring together that satisfies the condition that the most recent packet occurs within a threshold of time, the \textit{burst threshold}, of the previous packet. In other words, packets are grouped temporally and a new group is created only when no new packets have arrived within the amount of time set as the burst threshold. This is visually depicted in the \textit{Traffic Burstification} section of Fig.~\ref{fig_combined_outline}, where we can see Burst A and Burst B separated by the \textit{burst threshold}. We use the concept of a burst to logically divide the network traffic into discrete, manageable portions, which can then be further processed.

\textbf{Flow:} A flow is a sequence of packets (within a burst) with the same destination IP address and port number. That is, within a flow, all packets will either be going to (or coming from) the same destination IP address/port. Flows are not to be confused with TCP sessions. A flow ends at the end of a burst, while a TCP session can span multiple bursts. Thus, flows typically last for a few seconds, while TCP sessions can continue indefinitely. AppScanner leverages flows instead of TCP sessions to achieve real-time/near-to-real-time classification. From the \textit{Flow Separation} section of Fig.~\ref{fig_combined_outline}, it can be seen that a burst may contain one or more flows. Flows may overlap in a burst if a single app, \textit{App X}, initiates TCP sessions in quick succession or if another app, \textit{App Y}, happens to initiate a TCP session at the same time as \textit{App X}.\\

Our app identification framework first elicits network traffic from an app, generates features from that traffic, trains classifiers using these features, and finally identifies apps when the classifiers are later presented with unknown traffic.

\begin{figure*}[!ht]
\centering
\includegraphics[width=7.15in, clip=true, trim= 0.38in 6.1in 4.25in 0.52in]{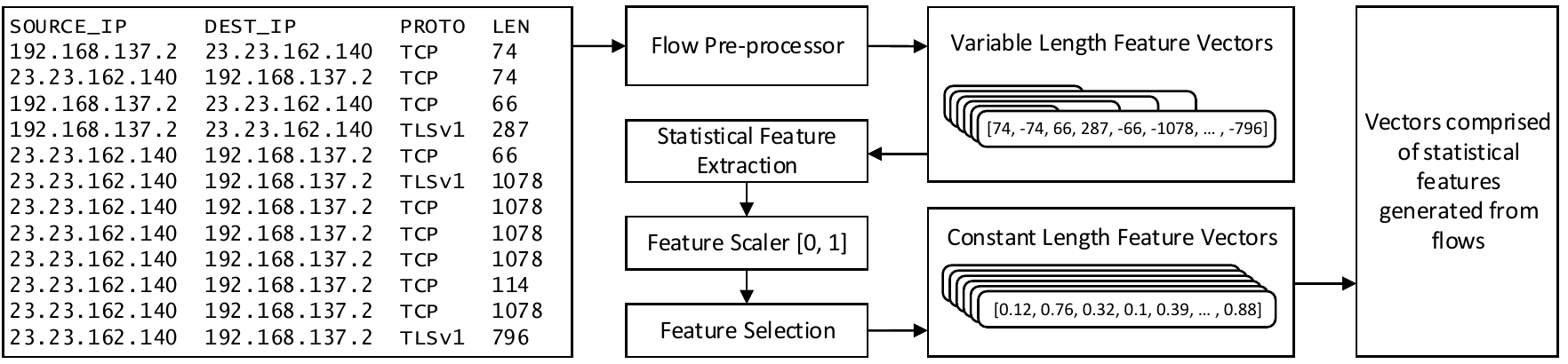}
\caption{Generating features from flows for classifier training.}
\label{fig_feature_generation}
\end{figure*}

\subsection{Equipment Setup}
\label{hardware_network_configuration}

The setup used to collect network traces from apps is shown in the \textit{Equipment Setup} section of Fig. \ref{fig_combined_outline}. The workstation was configured to forward traffic between the Wi-Fi access point (AP) and the Internet. To generate traffic from which to capture our training data, we used scripts that communicated with the target smartphone via USB using the Android Debug Bridge (ADB). These scripts were used to simulate user actions within apps and thus elicit network flows from the apps. This technique is called UI fuzzing.

The traffic generated by the smartphone was captured and exported as network traffic dumps containing details of captured packets. We collected packet details such as time, source address, destination address, ports, packet size, protocol and TCP/IP flags. The payload for each packet was also collected but was not used to provide features since it may or may not be encrypted. Although physical hardware was used for network traffic generation and capturing, this process can be massively automated and parallelized by running apps within Android emulators on virtual machines.

\subsection{Fingerprint Making}

There are several stages in the fingerprint making process as follows:\\

\noindent{\textbf{Network Trace Capture:} During traffic capture, we performed UI fuzzing on one app at a time to minimise `noise' (i.e., traffic generated simultaneously by other apps) in the network traces. Traffic from other apps or the Android operating system itself could interfere with and taint the fingerprint making process. To combat the problem of noise, the Network Log tool~\cite{networklog} was used to identify the app responsible for each network flow. Using data from Network Log combined with a `demultiplexing' script, all traffic that did not originate from the target app was removed from the traffic dump for that app. In this way, and in contrast to related work, we obtained perfect ground truth of what flows came from what app.

After data collection, the network traffic dumps were filtered to include only TCP traffic that was error free. For example, we filtered to remove packet retransmissions that were as a result of network errors.\\

\noindent{\textbf{Traffic Burstification and Flow Separation:} The next step was to parse the network dumps to obtain network traffic bursts. Traffic was first discretized into bursts to obtain ephemeral chunks of network traffic that could be sent immediately to the next stage of AppScanner for processing. This allows us to meet the design objective of real-time or near real-time classification of network traffic. Falaki et al. \cite{falaki2010first} observed that 95\% of packets on smartphones ``are received or transmitted within 4.5 seconds of the previous packet". During our tests, we observed that setting the burst threshold to one second instead of 4.5 seconds only slightly increased the number of bursts seen in the network traces. This suggests that network performance (in terms of bandwidth and latency) has improved since the original study. For this reason, we opted to use a burst threshold of one second to favour more overall bursts and nearer-to-real-time performance. Bursts were separated into individual flows (as defined at the beginning of this section and depicted in Fig. \ref{fig_feature_generation}) using destination IP address/port information. We enforced a maximum flow length that would be considered by the system. This is simply to ensure that abnormal traffic can be safely ignored in the real-world.}

It is important to note that while destination IP addresses were used for flow separation, they were not leveraged to assist with app identification. We also opted to not use information gleaned from DNS queries or flows with unencrypted payloads. We took this design decision to avoid the reliance on domain-specific knowledge that frequently changes, thus making our framework useful in the long term. Concretely, it is ill-advised to rely on the aforementioned additional sources of information for the following reasons:\\

\begin{itemize}
\item \textbf{IP addresses} - Destination IP addresses contacted by an app can change if DNS-based load-balancing/high-availability is used. Additionally, many apps communicate with similar IP addresses because they utilise the same CDN or belong to the same developer.
\item \textbf{DNS queries} - DNS queries are not always sent/observed due to the use of client-side DNS caching. Also, multiple apps may send the same DNS queries, for example, to resolve advertisement server domain names.
\item \textbf{Packet payloads} - Many app developers are becoming more privacy-aware and are opting to use SSL/TLS to encrypt packet payloads. Thus features extracted from TCP payloads will become less useful over time.\\
\end{itemize}

\noindent{\textbf{Ambiguity Detection:} As mentioned at the beginning of this section, many apps have third-party libraries in common (especially ad libraries) and these libraries themselves generate network traffic. Unfortunately, it is not possible to discriminate traffic coming from libraries (as opposed to the app that embeds the library) in a scalable way, i.e., without an intrusive approach such as reverse-engineering or modifying apps. Indeed, as far as the operating system is concerned, apps and their bundled libraries are one entity within the same process. Since network traffic generated by libraries in common across apps is similar, this will frustrate the fingerprinting process because classifiers will be given contradictory training examples. This problem of so-called \textit{ambiguous flows} poses a challenge to naive machine learning approaches. To mitigate negative effects, we introduce \textit{Ambiguity Detection} as detailed in Section~{\ref{section_ambiguity_detection}. Ambiguity detection uses simple reinforcement learning techniques to identify similar flows coming from different apps. In the training phase, ambiguous flows are detected and relabelled as belonging to the ``ambiguous" class, so that the system is later able to properly identify and handle them.\\}

\noindent{\textbf{Classifier Training:} Statistical features were generated from flows and used to train classifiers. Statistical feature extraction involves deriving 54 statistical features from each flow as shown in Figure~\ref{fig_feature_generation}. For each flow, three packet series are considered: incoming packets only, outgoing packets only, and bi-directional traffic (i.e. both incoming and outgoing packets). For each series (3 in total), the following values were computed: minimum, maximum, mean, median absolute deviation, standard deviation, variance, skew, kurtosis, percentiles (from 10\% to 90\%) and the number of elements in the series (18 in total). These statistical features are computed using the Python pandas \cite{mckinney-proc-scipy-2010} libraries.

These features are then passed through the Feature Scaler function, which is a min-max scaler (i.e., the minimum and the maximum value for a specific feature in the training set corresponds to 0 and 1 respectively). In order to avoid the curse of dimensionality, the Feature Selection function is used to choose the best features. Feature Selection relies on the significance score given to each feature by the estimators of a Random Forest classifier that was run on the training set. At this point, we selected only those features with a score higher than 1\%, for a total of 40 features of the original 54.\\}

\subsection{App Identification}

Unknown flows are passed to the trained classifiers. Ambiguous flows are identified and labelled as such, since the classifiers were trained to understand ambiguous flows. Flows that are not labelled by the classifiers as ambiguous next go through \textit{classification validation} as described in Section~\ref{section_classification_validation}. The classification validation stage is crucial for one primary reason. Machine learning algorithms will always attempt to place an unlabelled example into the class it most closely resembles, even if the match is not very good. Given that our classifiers will never be trained with the universe of flows from apps, it follows that there will be some flows presented to AppScanner which are simply unknown or never-before-seen. If left unchecked, this can cause an undesirable increase in the false positive (FP) rate.

To counteract these problems, we leverage the prediction probability metric (available in many classifiers) to understand how certain the classifier is about each of its classifications. For example, if the classifier labelled an unknown sample as \textit{com.facebook.katana}, we would check its prediction probability value for that classification to determine the classifier's confidence. If this value is below the classification validation threshold, AppScanner will not make a pronouncement. However, if this value exceeds the threshold, AppScanner would report it as a match for that particular app. In Section \ref{section_improving_accuracy}, we discuss how varying this threshold impacts the precision, recall, and overall accuracy of AppScanner, as well as how this affects the percentage of total flows that the classifiers are confident enough to classify.

\begin{comment}
\begin{figure}[!ht]
\centering
\includegraphics[width=3.5in, clip=true, trim= 0.38in 6.1in 7.5in 0.35in]{flow-matching.pdf}
\caption{High-level representation of identifying unknown flows. Unknown flows first pass through the ambiguity detection phase.}
\label{fig_flow_identification}
\end{figure}
\end{comment}
\section{Ambiguity Detection}
\label{section_ambiguity_detection}

The ambiguity detection phase aims to identify and relabel ambiguous flows. This phase involves a reinforcement learning strategy that is leveraged during classifier training. As outlined in Fig.~\ref{fig_reinforcement_learning}, classifier training is divided in two stages: the \textit{preliminary classifier} stage, and the \textit{reinforced classifier} stage.

\begin{figure*}[!ht]
\centering
\includegraphics[width=7.15in, clip=true, trim= 0.38in 6.5in 4.25in 0.52in]{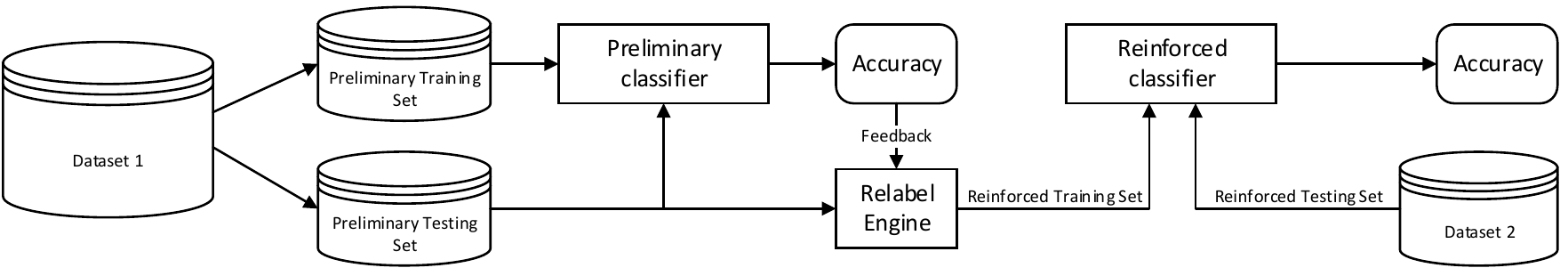}
\caption{Using reinforcement learning to obtain robustness against ambiguous flows.}
\label{fig_reinforcement_learning}
\end{figure*}

The main training set considered in the analysis is first randomly shuffled and divided into halves: the \textit{preliminary training set} and the \textit{preliminary testing set}. The preliminary training set is used to train the preliminary classifier. The preliminary testing set is used to measure the accuracy of the preliminary classifier, and as a basis for generating the training set for the reinforced classifier. In this way, we can first identify which flows are incorrectly classified by the preliminary classifier. We validated that these incorrectly labelled flows are to a large extent library traffic, as expected.

The Relabel Engine leverages feedback on the accuracy of the preliminary classifier to identify ambiguous flows. Flows in the preliminary testing set that are incorrectly classified are re-labelled as ``ambiguous" by the \textit{Relabel Engine}. On the other hand, flows that are correctly classified by the preliminary classifier keep their original label (i.e., the app that generated them). This relabelled dataset is now used as the reinforced training set and is passed to the reinforced classifier. The reinforced classifier is thus equipped to identify ambiguous flows since it is trained with examples of ambiguous flows.

We emphasise to the reader that no flows from the preliminary training set are used in the reinforced training set. The preliminary classifier and the preliminary training set are only used as a means of identifying ambiguous flows so that additional knowledge can be provided to the reinforced classifier.

\section{Dataset Collection}

To test the performance of AppScanner, we considered a random 110 of the 200 most popular free apps as listed by the Google Play Store. We chose the most popular apps because they form a large part of the install-base of apps across the world. Additionally, we chose free apps because free apps tend to be ad-supported and thus use ad libraries. There is a small set of major ad libraries and thus ad libraries tend to be shared across apps. This suggests that free apps will be more likely to generate ambiguous flows than paid apps. Being able to properly fingerprint and identify free apps thus implies that AppScanner is robust enough to handle paid apps as well.

Smartphones in our testbed were connected to the internet via a Linksys E1700 Wi-Fi router/AP that had its internet connection routed through a workstation. UI fuzzing was performed on each app for~30 minutes. UI fuzzing simulated user actions by invoking UI events such as touches, swipes, and button presses. These UI events were generated randomly and sent to apps. It is worth noting that some apps presented login screens upon first launch. In those cases, we first manually created accounts for those apps before logging in. We did this to ensure that traffic generation using UI fuzzing was not hindered by a login screen. Greater coverage of all the network flows in an app may theoretically be obtained by using advanced UI fuzzing techniques provided by frameworks such as Dynodroid~\cite{machiry2013dynodroid}, or by recruiting human participants. However, we consider these approaches to be out of the scope of our research.

\subsection{Dataset Collection}

A major contribution of this work is to understand how app fingerprinting is affected by time, the device used, app versions, and combinations of these variables. For this reason, we collected several datasets as outlined in Table~\ref{table-experiment-setup}. In what follows, we describe these datasets in detail.

The dataset we consider as our baseline is \texttt{Dataset-1}, which was collected using \textit{Device-A}, a Motorola XT1039 with Android version 4.4.4 as the operating system. This dataset contains network traffic from 110 apps that were the latest versions of each app at the time of initial data collection. We refer to this time of initial data collection as $T_0$. All other datasets (\texttt{Dataset-2} to \texttt{Dataset-5}) were collected six months after $T_0$, i.e., at time $T_0$ + 6 months.

\texttt{Dataset-2} differs from \texttt{Dataset-1} only by the time of data collection. \texttt{Dataset-2} contains data from only 65 apps (instead of 110), because the remaining 45 apps refused to run without being updated. We hereafter refer to the 65 apps in \texttt{Dataset-2} that ran without being updated as the \textit{run-without-update} subset. 

\texttt{Dataset-3} was collected using \textit{Device-B}, an LG E960 with Android version 5.1.1 as the operating system. \texttt{Dataset-3} also used the \textit{run-without-update} subset.

\texttt{Dataset-4} and \texttt{Dataset-5} were obtained by collecting network traffic from the latest versions (at the time of data collection six months after initial data collection) of the original 110 apps and were collected using \textit{Device-A} and \textit{Device-B} respectively.

Additionally, we consider variants of the aforementioned datasets, which consider only apps in the \textit{run-without-update} subset. We denote these dataset variants as \texttt{Dataset-1a}, \texttt{Dataset-4a}, and \texttt{Dataset-5a} for \texttt{Dataset-1}, \texttt{Dataset-4} and \texttt{Dataset-5}, respectively. These datasets were generated in order to offer a balanced analysis in the presence of datasets with different numbers of apps (i.e., \texttt{Dataset-2} and \texttt{Dataset-3}).

\begin{table*}[!th]
\centering
\caption{Descriptions of the devices, operating systems, number of apps, app versions, and time of data collection for each dataset used.}
\label{table-experiment-setup}
\begin{tabular}{@{}p{0.87in}lccp{1.8in}l@{}}
\hline
\textbf{Name} & \textbf{Device} & \textbf{Operating System} & \textbf{Number of apps} & \textbf{App versions} & \textbf{Time of data collection} \\ \hline
\texttt{Dataset-1}  & Motorola XT1039 & Android 4.4.4 & 110 & Latest versions as at $T_0$            & $T_0$            \\ \hline
\texttt{Dataset-1a} & Motorola XT1039 & Android 4.4.4 & 65  & Latest versions as at $T_0$            & $T_0$            \\ \hline
\texttt{Dataset-2}  & Motorola XT1039 & Android 4.4.4 & 65  & Latest versions as at $T_0$            & $T_0$ + 6~months \\ \hline
\texttt{Dataset-3}  & LG E960         & Android 5.1.1 & 65  & Latest versions as at $T_0$            & $T_0$ + 6~months \\ \hline
\texttt{Dataset-4}  & Motorola XT1039 & Android 4.4.4 & 110 & Latest versions as at $T_0$ + 6~months & $T_0$ + 6~months \\ \hline
\texttt{Dataset-4a} & Motorola XT1039 & Android 4.4.4 & 65  & Latest versions as at $T_0$ + 6~months & $T_0$ + 6~months \\ \hline
\texttt{Dataset-5}  & LG E960         & Android 5.1.1 & 110 & Latest versions as at $T_0$ + 6~months & $T_0$ + 6~months \\ \hline
\texttt{Dataset-5a} & LG E960         & Android 5.1.1 & 65  & Latest versions as at $T_0$ + 6~months & $T_0$ + 6~months \\ \hline
\end{tabular}
\end{table*}

\section{Evaluation}
\label{section_evaluation}

In evaluating our system, we followed a number of steps. First, we report the results of a baseline evaluation of system performance using training and testing sets derived from single datasets. Second, to obtain a more representative measurement of system performance, we performed a comprehensive suite of tests (as outlined in Table~\ref{table_various_tests}) using completely independent training and testing sets. Measurements were taken to understand how factors such as time, device (including operating system), app version, and a combination of device and app version affected performance. 

We leveraged the \textit{scikit-learn}~\cite{sklearn_api} machine learning libraries to implement the classifiers in our framework. Random Forest classifiers were chosen since they gave superior performance in our previous work. All classifiers were set to use default parameters.

We highlight to the reader that any results reported in this section should be considered as lower bounds of system performance. Indeed, the results presented in this section show the performance of the system before any performance-enhancers, such as ambiguity detection and classification validation (Section~\ref{section_improving_accuracy}), have been applied. The tests performed in this section are merely to assess default system performance before post-processing is applied.

For our baseline results, we split each dataset into a training set (75\% of examples) and a testing set (25\% of examples) and used them to train classifiers as detailed in Section~\ref{system-design}. The performance of these classifiers is shown in Table~\ref{table_baseline}. Accuracy within datasets fell between 66.4\% and 73.5\%. We underscore that these results are obtained without applying any post-processing. These results are fairly good but may overestimate the performance of the system. This is because the training and testing sets in each case were generated from one original dataset. In what follows, we do more robust measurements by using completely independent datasets  for training and testing to make a more real-world assessment of system performance.

\subsection{Effect of Time}

To measure the effect of time on classification performance, we trained a classifier with \texttt{Dataset-1a} and tested with \texttt{Dataset-2}. This combination of training and testing sets assessed the effect of keeping device and app versions constant, but causing six months to pass between collection of data for training and testing. The overall accuracy for this test, called the \texttt{TIME} test, was 40.5\% and had the highest performance of our tests that used completely separate training and testing sets.

The fact that this test gave the highest performance of tests with completely independent training and testing sets is expected, since the app versions and device (including operating system) were constant. That is, the logic (app and operating system) that generates traffic seems to generate the same traffic even after some amount of time (in this case six months) has elapsed. Since the underlying logic does not change, it would be reasonable to expect app fingerprints to also remain constant.

\subsection{Effect of a Different Device}

To assess the impact of a different device on app classification we did three tests: \texttt{D-110}, \texttt{D-110A}, and \texttt{D-65}. \texttt{D-110} used \texttt{Dataset-4} as a training set and \texttt{Dataset-5} as a testing set. That is, we trained with 110 apps on one device and tested with the same 110 apps on a different device. The overall accuracy was 37.2\%. \texttt{D-110A} used the \textit{run-without-update} subsets of the datasets used in \texttt{D-110} and had an overall accuracy of 37.5\%. \texttt{D-65} consisted of a training set of \texttt{Dataset-2} and testing set of \texttt{Dataset-3}. That is, we trained with 65 apps on one device and tested with 65 apps on another device. The overall accuracy for this test was 39.0\%. We note that this test, with 65 apps, gives similar performance to the \texttt{TIME} test, which also had 65 apps. This insight suggests that device model and operating system version does not have a major effect on app fingerprinting performance.

\begin{table}[!t]
\centering
\caption{Baseline performance of app classification for each dataset without any post-processing techniques applied.}
\label{table_baseline}
\begin{tabular}{@{}lcccc@{}}
\hline
\textbf{Dataset} & \textbf{Precision(\%)} & \textbf{Recall(\%)} & \textbf{F1(\%)} & \textbf{Accuracy (\%)} \\ \hline
\texttt{Dataset-1} & 74.0 & 71.6 & 72.0 & 72.5 \\ \hline
\texttt{Dataset-1a} & 74.0 & 72.9 & 72.9 & 73.5 \\ \hline
\texttt{Dataset-2} & 68.3 & 68.1 & 67.7 & 66.4 \\ \hline
\texttt{Dataset-3} & 71.3 & 68.8 & 69.5 & 69.7 \\ \hline
\texttt{Dataset-4} & 68.1 & 66.3 & 66.5 & 67.3 \\ \hline
\texttt{Dataset-4a} & 68.2 & 66.7 & 66.9 & 66.7 \\ \hline
\texttt{Dataset-5} & 69.5 & 68.2 & 68.3 & 69.6 \\ \hline
\texttt{Dataset-5a} & 70.2 & 67.0 & 67.6 & 68.2 \\ \hline
\end{tabular}
\end{table}

\begin{table*}[!t]
\centering
\caption{Summary of the comprehensive suite of tests used to measure the performance of the app classification system. All training and testing sets were completely independent of each other. The independent variables for each test are identified.}
\label{table_various_tests}
\begin{tabular}{@{}lllccccp{1.5in}c@{}}
\hline
\textbf{Test Name} & \textbf{Training Set} & \textbf{Testing Set} & \textbf{Precision (\%)} & \textbf{Recall (\%)} & \textbf{F1 (\%)} & \textbf{Accuracy (\%)} & \textbf{Independent Variable} & \textbf{Apps} \\ \hline

\texttt{TIME}   & \texttt{Dataset-1a} & \texttt{Dataset-2}  & 44.2 & 43.0 & 42.3 & 40.5 & Time                   & 65  \\ \hline
\texttt{D-110}  & \texttt{Dataset-4}  & \texttt{Dataset-5}  & 40.3 & 36.0 & 35.7 & 37.2 & Device                 & 110 \\ \hline
\texttt{D-110A} & \texttt{Dataset-4a} & \texttt{Dataset-5a} & 38.7 & 34.9 & 35.0 & 37.6 & Device                 & 65 \\ \hline
\texttt{D-65}   & \texttt{Dataset-2}  & \texttt{Dataset-3}  & 43.5 & 38.0 & 38.7 & 39.0 & Device                 & 65  \\ \hline
\texttt{V-LG}   & \texttt{Dataset-3}  & \texttt{Dataset-5a} & 32.8 & 31.2 & 30.2 & 30.4 & App versions           & 65  \\ \hline
\texttt{V-MG}   & \texttt{Dataset-2}  & \texttt{Dataset-4a} & 34.8 & 32.1 & 32.3 & 32.8 & App versions           & 65  \\ \hline
\texttt{DV-110} & \texttt{Dataset-1}  & \texttt{Dataset-5}  & 23.7 & 19.5 & 19.5 & 19.4 & Device \& App versions & 110 \\ \hline
\texttt{DV-65}  & \texttt{Dataset-1a} & \texttt{Dataset-5a} & 20.4 & 19.3 & 18.4 & 19.0 & Device \& App versions & 65  \\ \hline
\end{tabular}
\end{table*}

\subsection{Effect of Different App Versions}

We carried out two tests to understand the impact that different app versions had on app fingerprinting. \texttt{V-LG} involved training with \texttt{Dataset-3} and testing with \texttt{Dataset-5a}. For this test, the same device was used but with different versions of the same apps. The overall accuracy of this test was 30.4\%. \texttt{V-MG} used a training set of \texttt{Dataset-2} and testing set of \texttt{Dataset-4a}. The overall accuracy for this test was 32.8\%. We note that the accuracy for both of these tests were fairly similar but markedly lower than the \texttt{TIME}, \texttt{D-110} or \texttt{D-110A} or \texttt{D-65} tests. This insight suggests that changes in app versions affects the reliability of app fingerprinting. We believe that this phenomenon could be due to changes in app code or logic that has direct consequences on the way that an app generates network flows. Thus there is a need to keep app fingerprint databases up-to-date as app developers release new app versions.

\subsection{Effect of a Different Device and Different App Versions}

A final two tests were conducted to measure the impact of changing both device and app versions. The first test, \texttt{DV-110}, used a training set of \texttt{Dataset-1} and a testing set of \texttt{Dataset-5}, i.e., using a total of 110 apps. The second test, \texttt{DV-65}, used a training set of \texttt{Dataset-1a} and testing set of \texttt{Dataset-5a}. These tests yielded overall accuracies of 19.4\% and 19.0\% respectively. As expected from the results of our previous tests, changing both device and app versions together more severely impacted classification performance. It is interesting to note, however, that the number of apps in training and testing sets did not seem to impact overall classification accuracy in a negative way under these adverse conditions. This lends support to the idea that app fingerprinting can be a scalable process. We note that despite \texttt{DV-110} and \texttt{DV-65} having approximately half the accuracy of the \texttt{TIME} test, they still perform approximately 20 times better than pure random guessing.
\section{Improving Accuracy}
\label{section_improving_accuracy}

Our results so far show the performance of AppScanner without any post-processing applied. In this section, we look at two post-processing strategies that have proven effective in improving the accuracy of the system: ambiguity detection and classification validation. Ambiguity detection is detailed in Section~\ref{section_ambiguous_detection} and classification validation is discussed in Section~\ref{section_classification_validation}. In general, both of these strategies aim to identify network flows that are not reliable for app fingerprinting.

\subsection{Ambiguity Detection}
\label{section_ambiguous_detection}

As mentioned in Section~\ref{section_ambiguity_detection}, many apps have traffic in common and this can hinder app classification if left unhandled. Our reinforcement learning approach identifies and relabels ambiguous flows so that the classifiers have a model to identify them. When measuring performance with ambiguity detection in use, unknown flows that are labelled as ambiguous are omitted from calculations of classifier performance. That is, ambiguous flows are identified and ignored, and thus do not affect the measurement of the performance of our system. This was done to ensure that measurements of system performance were not artificially inflated when the classifiers correctly identified ambiguous flows.

In what follows, we report on the improvements that can be made by using our reinforcement learning approach to identify ambiguous traffic flows. Table~\ref{table_reinforcement_beforeafter} shows the improvement in performance obtained by applying ambiguity detection as outlined in Figure~\ref{fig_reinforcement_learning}. Each test uses the same training and testing sets as described in Table~\ref{table_various_tests}, with the only change being that reinforced classifiers are used instead. Ambiguity detection was applied to the training sets of these reinforced classifiers as detailed in Section~\ref{section_ambiguity_detection}.

Reinforced classifiers received a boost in overall accuracy of approximately 1.5-2.1 times. Precision, recall, and F-1 score saw similar increases. The most challenging tests, \texttt{DV-110} and \texttt{DV-65} (using different physical devices, Android versions, and app versions between training and testing sets), had the greatest percentage increases in performance and saw accuracy approximately double when using reinforced classifiers. For example, in \texttt{DV-110}, accuracy was increased from 19.4\% to 41.0\% using ambiguity detection. Improving performance using reinforced classifiers highlights the prevalence of ambiguous flows in app traffic and reiterates the need for systems that can address them.

\begin{table*}[]
\centering
\caption{How the reinforcement learning strategy improved classifier performance for each of the tests that were conducted.}
\label{table_reinforcement_beforeafter}
\begin{tabular}{@{}p{1.3in}|cccc|cccc@{}}
\hline
\multirow{2}{*}{\textbf{Test Name}} & \multicolumn{4}{c|}{\textbf{Preliminary Classifier}} & \multicolumn{4}{c}{\textbf{Reinforced Classifier}} \\
 & \textbf{Precision (\%)} & \textbf{Recall (\%)} & \textbf{F-1 (\%)} & \textbf{Accuracy (\%)} & \textbf{Precision (\%)} & \textbf{Recall(\%)} & \textbf{F-1 (\%)} & \textbf{Accuracy (\%)} \\ \hline
\texttt{TIME} & 44.2 & 43.0 & 42.3 & 40.5 & 66.9 & 65.7 & 65.2 & 72.2 \\ \hline
\texttt{D-110} & 40.3 & 36.0 & 35.7 & 37.2 & 59.9 & 56.2 & 55.8 & 66.3 \\ \hline
\texttt{D-110A} & 38.7 & 34.9 & 35.0 & 37.6 & 57.6 & 52.4 & 52.6 & 62.6 \\ \hline
\texttt{D-65} & 43.5 & 38.0 & 38.7 & 39.0 & 62.4 & 58.5 & 57.8 & 65.5 \\ \hline
\texttt{V-LG} & 32.8 & 31.2 & 30.2 & 30.4 & 46.5 & 44.3 & 42.9 & 51.1 \\ \hline
\texttt{V-MG} & 34.8 & 32.1 & 32.3 & 32.8 & 50.9 & 49.4 & 48.4 & 57.2 \\ \hline
\texttt{DV-110} & 23.7 & 19.5 & 19.5 & 19.4 & 38.0 & 35.4 & 33.9 & 41.0 \\ \hline
\texttt{DV-65} & 20.4 & 19.3 & 18.4 & 19.0 & 36.2 & 33.8 & 31.7 & 37.2 \\ \hline
\end{tabular}
\end{table*}

\subsection{Classification Validation}
\label{section_classification_validation}

Classification validation is another effective strategy that can be leveraged to improve app classification performance. Classifiers can be made to output their confidence when labelling and unknown example. In simple terms, a classifier may be very confident about a classification if the class boundaries within its models are distinct, i.e., with sufficient separation between classes. In other cases, this distinction may be less clear.

By assessing the confidence that a classifier reports with its classification, a judgement can be made as to whether the classification will be considered as valid by the system. We call the cut-off for what is considered a valid classification the prediction probability threshold. A higher prediction probability threshold will lead to more conservative predictions, and thus higher accuracy, at the expense of the number of flows whose classification is considered as valid. On the other hand, a lower threshold reduces accuracy but maximises the number of flows whose classification is considered as valid. For app classification, false positives are usually undesirable and thus higher prediction probability thresholds are likely to be suitable.

Classification validation reduces the number of flows that are considered as being ``correctly" classified, but it is important to note that there is no inherent requirement to label all unknown flows. Apps typically send tens or hundreds of flows per minute when they are being used, so there remains significant opportunity to identify apps from their more distinctive flows. Thus, classification validation can be an effective technique for improving app classification performance while incurring negligible drawback. In what follows, we report on the improvements provided by classification validation to reinforced classifiers.

Figure~\ref{fig_reinforced_results} shows the improvement provided by classification validation for the \texttt{TIME}, \texttt{D-110}, \texttt{D-110A}, and \texttt{D-65} tests. We highlight some results by considering a prediction probability threshold of 0.9. Figure~\ref{fig_reinforced_time} shows that the \texttt{TIME} test had a preliminary accuracy of 72\% which was improved to 96\% using classification validation. The results for the \texttt{D-110} and \texttt{D110-A} tests are shown in Figures~\ref{fig_reinforced_d110} and \ref{fig_reinforced_d110a} respectively. Overall accuracy was improved from 66\% to 88\% for test \texttt{D-110} and from 63\% to 92\% for \texttt{D-110A}. It is interesting to note that \texttt{D-110A} had a higher peak accuracy than \texttt{D-110} after using classification validation, although \texttt{D-110} had a higher baseline accuracy. The final test in this group, \texttt{D-65} saw accuracy go from 66\% to 93\% when using classification validation.

\begin{figure*}[t!]
\centering
	\subfloat[Performance for the \texttt{TIME} test.
	\label{fig_reinforced_time}]
	{
		\includegraphics[width=3.3in,clip=true]{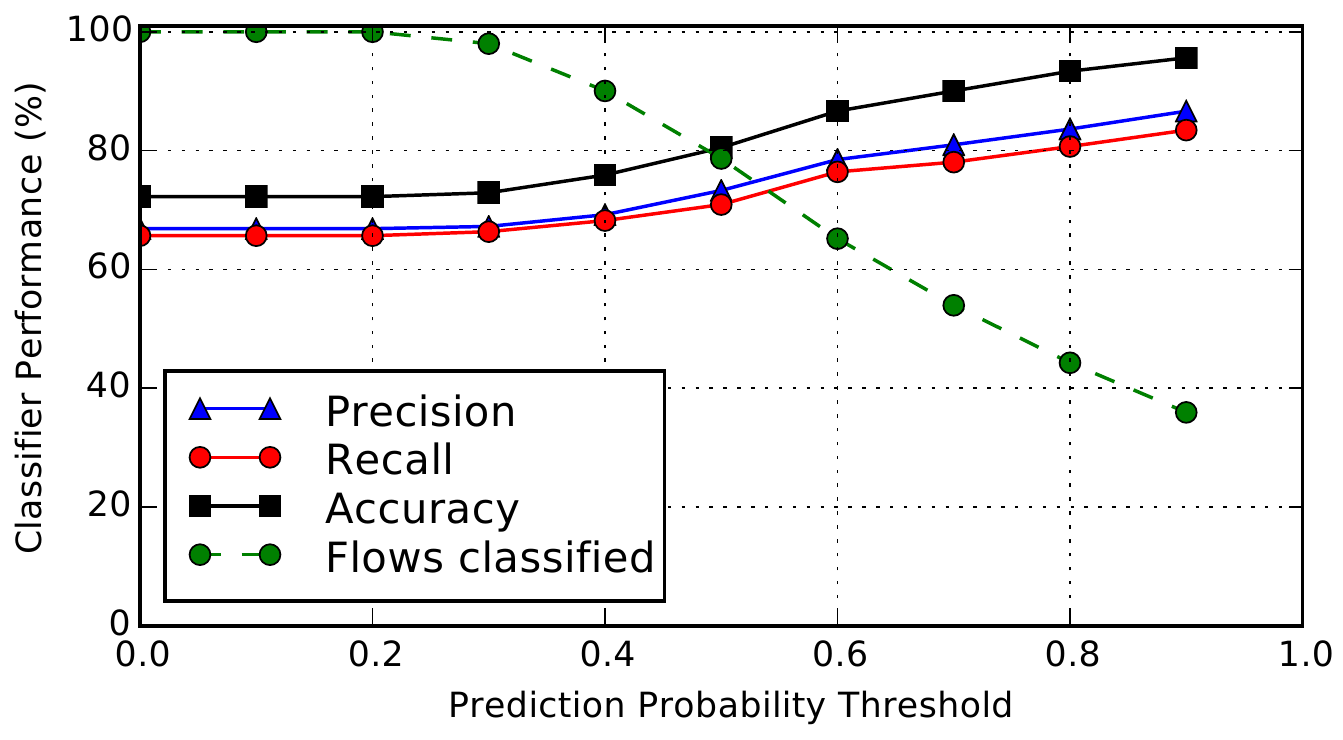}
    }
	\hspace{0.75cm}
	\subfloat[Performance for the \texttt{D-110} test.
	\label{fig_reinforced_d110}]
	{
		\includegraphics[width=3.3in,clip=true]{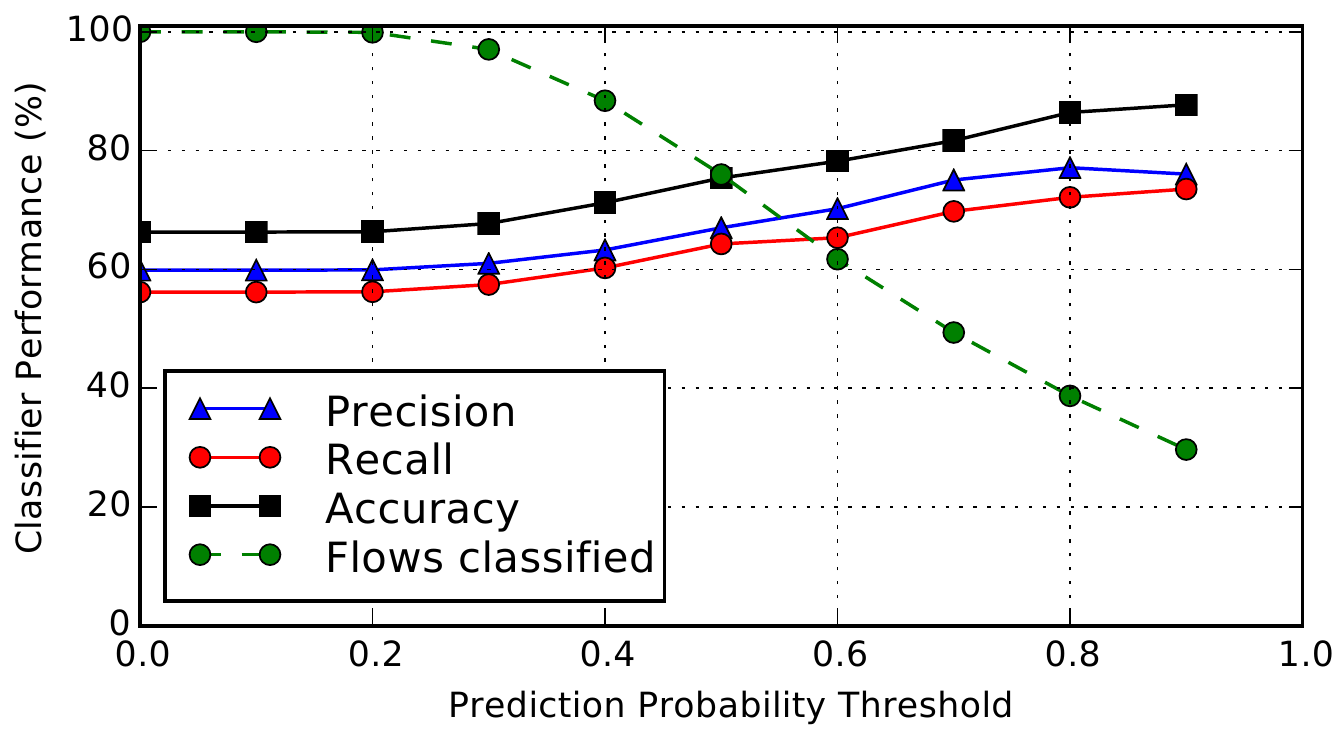}
	}
	\vspace{0.05cm}
	\subfloat[Performance for the \texttt{D-110A} test.
	\label{fig_reinforced_d110a}]
	{
		\includegraphics[width=3.3in,clip=true]{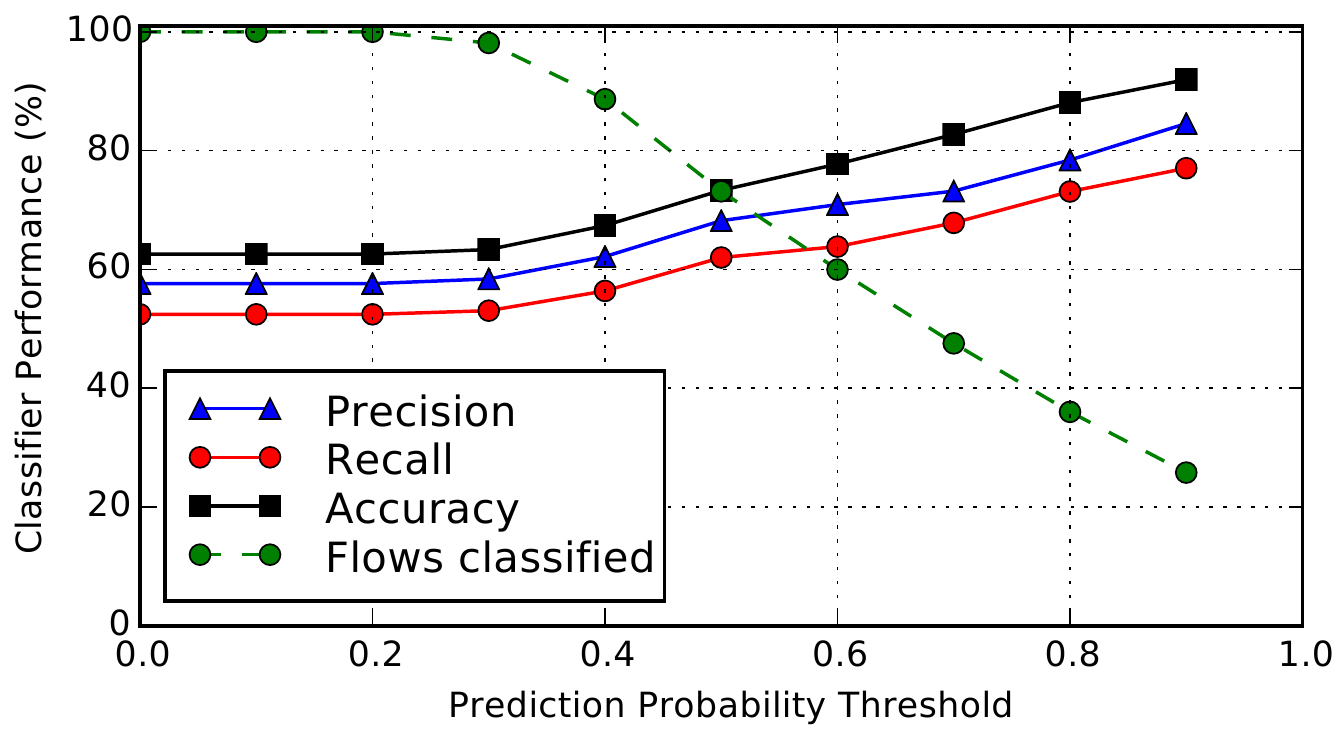}
    }
	\hspace{0.75cm}
	\subfloat[Performance for the \texttt{D-65} test.
	\label{fig_reinforced_d65}]
	{
		\includegraphics[width=3.3in,clip=true]{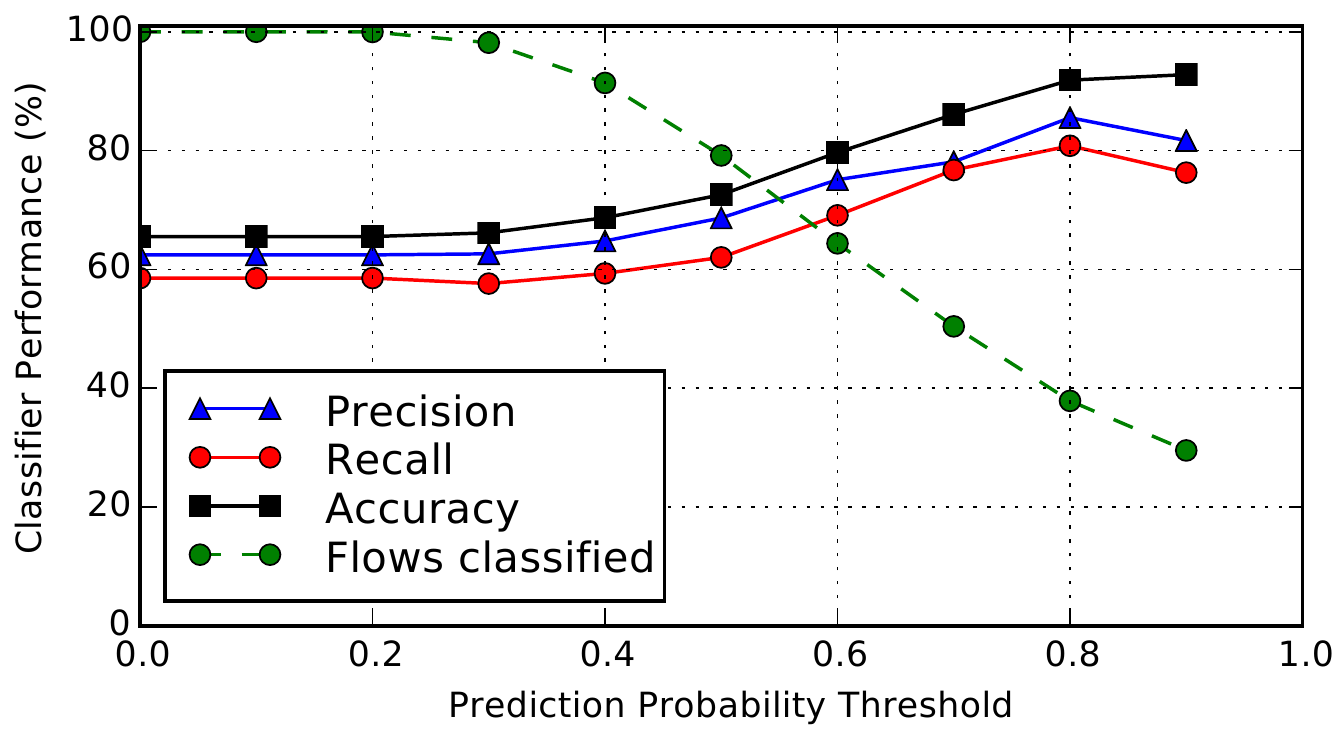}
	}
\caption{Performance of our reinforced classifiers on the \texttt{TIME}, \texttt{D-110}, \texttt{D-110A}, and \texttt{D-65} tests.}
\label{fig_reinforced_results}
\end{figure*}

Figure~\ref{fig_reinforced_results2} shows the improvement provided by classification validation for the \texttt{V-LG}, \texttt{V-MG}, \texttt{DV-110}, and \texttt{DV-65} tests. Once again, we report our results assuming a prediction probability threshold of 0.9 is chosen. Figure~\ref{fig_reinforced_vlg} shows that classification validation improved accuracy for the \texttt{V-LG} test from 51\% to 84\%. Figure~\ref{fig_reinforced_vmg} shows the results for the \texttt{V-MG} test, which is similar to \texttt{V-LG} but with a different device. Classification validation improved accuracy from 57\% to 79\% in this case.

Figures~\ref{fig_reinforced_dv110} and \ref{fig_reinforced_dv65} show the results for our most challenging tests: \texttt{DV-110} and \texttt{DV-65}. Classification validation was able to increase the accuracy of \texttt{DV-110} from 41\% to 73\%. Likewise, for test \texttt{DV-65}, accuracy was increased from 37\% to 76\%. This demonstrates that classification validation can be a useful tool to improve system performance under difficult conditions.

\begin{figure*}[t!]
\centering
	\subfloat[Performance for the \texttt{V-LG} test.
	\label{fig_reinforced_vlg}]
	{
		\includegraphics[width=3.3in,clip=true]{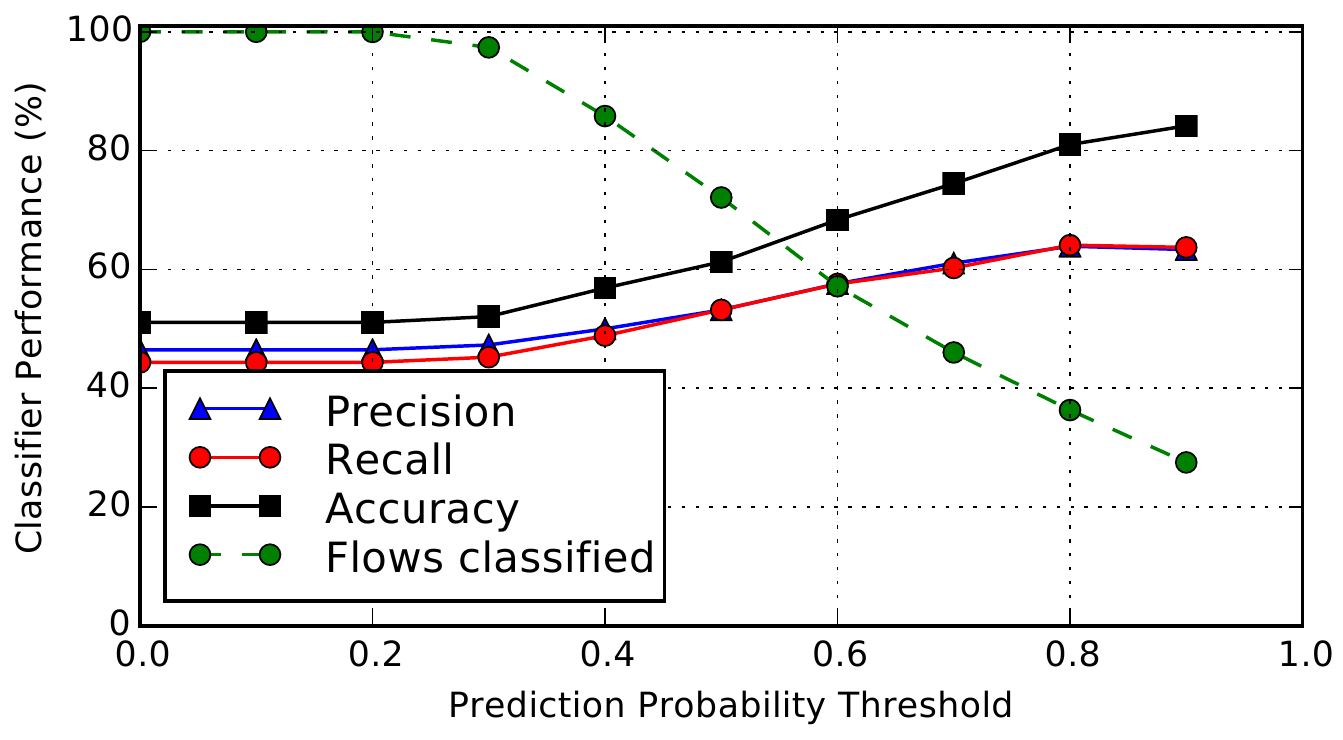}
    }
	\hspace{0.75cm}
	\subfloat[Performance for the \texttt{V-MG} test.
	\label{fig_reinforced_vmg}]
	{
		\includegraphics[width=3.3in,clip=true]{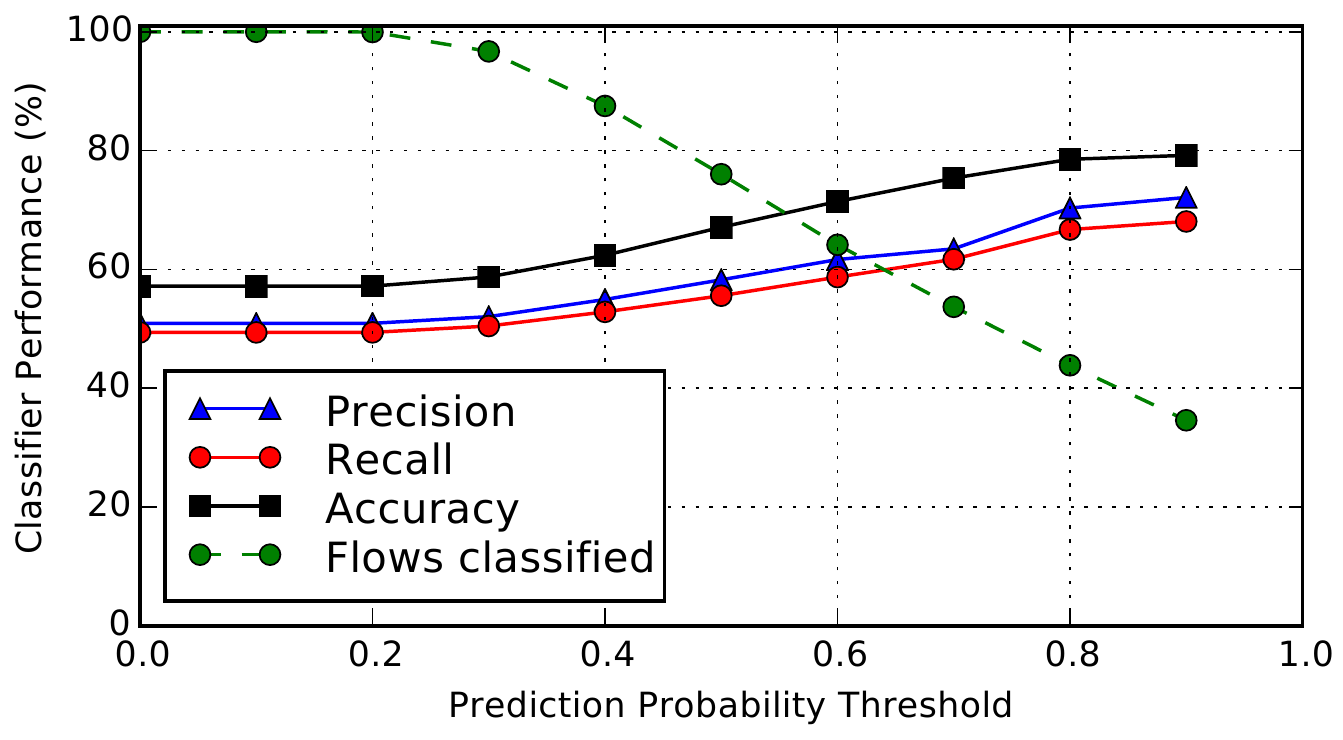}
	}
	\vspace{0.05cm}
	\subfloat[Performance for the \texttt{DV-110} test.
	\label{fig_reinforced_dv110}]
	{
		\includegraphics[width=3.3in,clip=true]{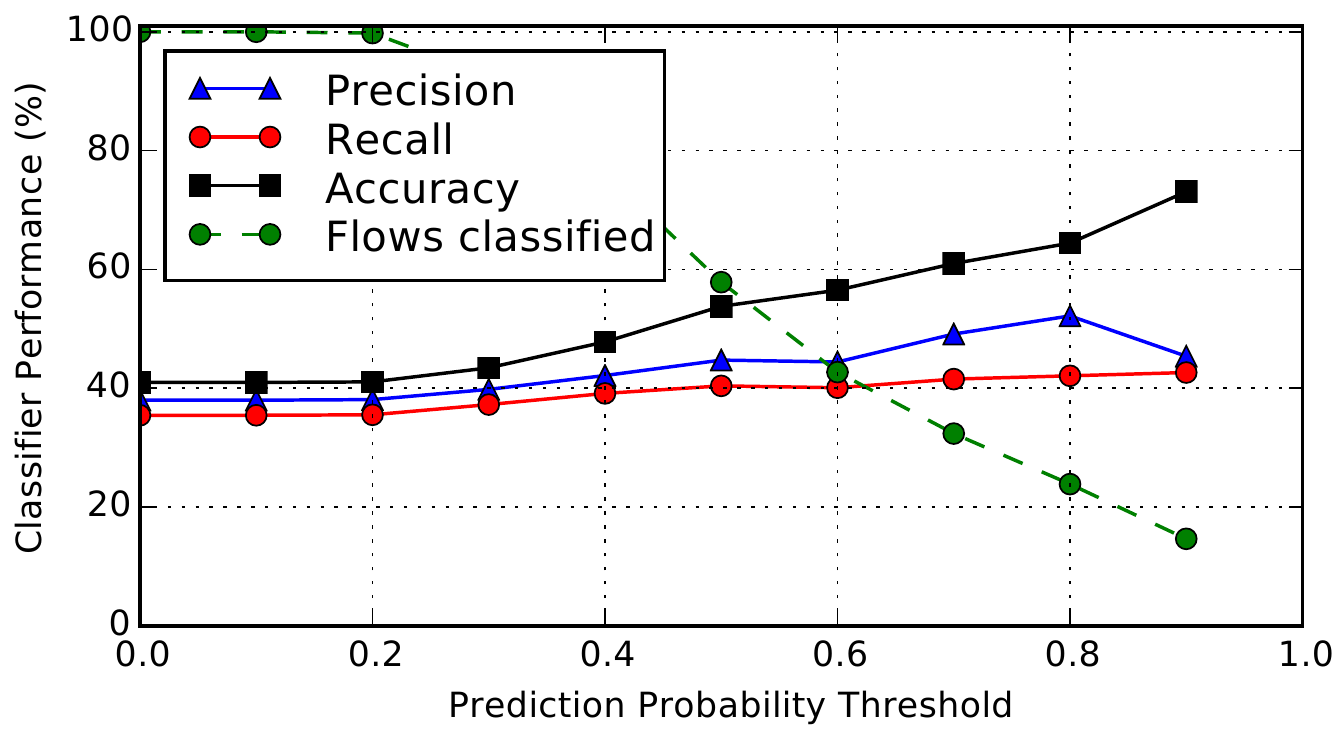}
    }
	\hspace{0.75cm}
	\subfloat[Performance for the \texttt{DV-65} test.
	\label{fig_reinforced_dv65}]
	{
		\includegraphics[width=3.3in,clip=true]{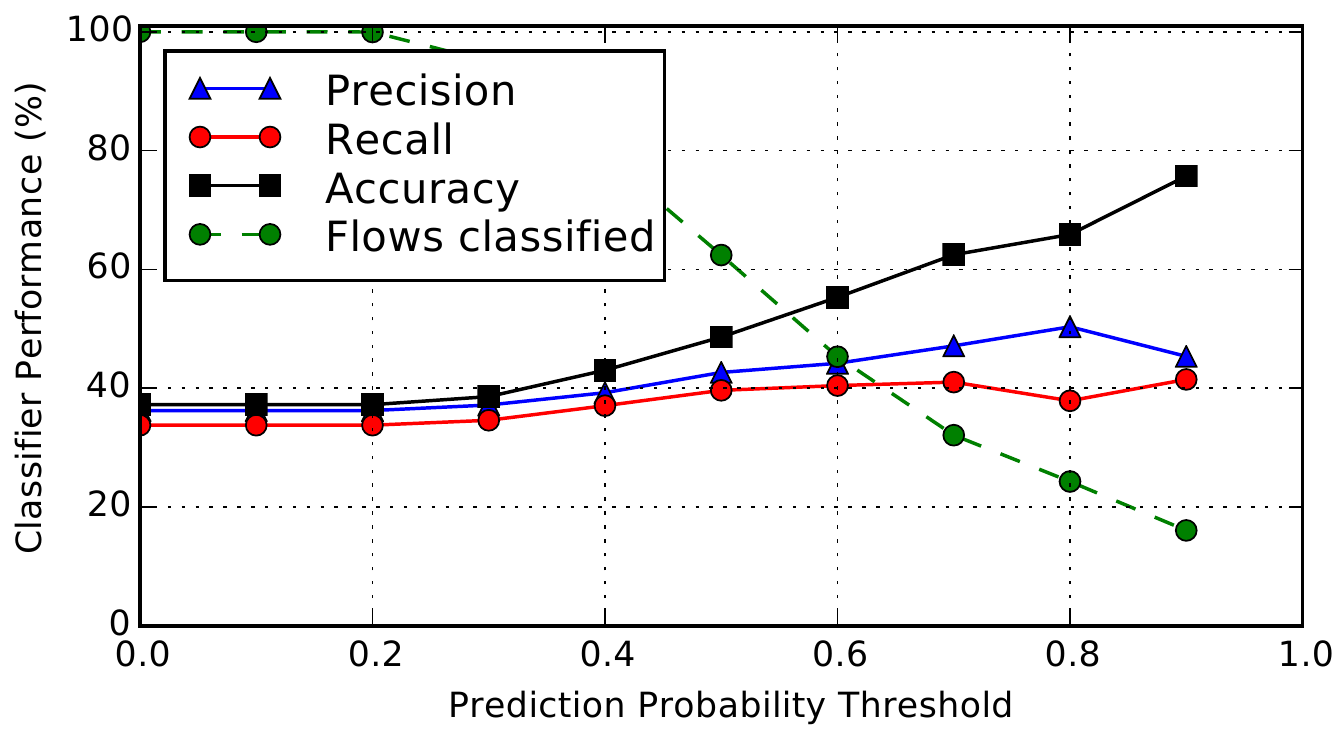}
	}
\caption{Performance of our reinforced classifiers on the \texttt{V-LG}, \texttt{V-MG}, \texttt{DV-110}, and \texttt{DV-65} tests.}
\label{fig_reinforced_results2}
\end{figure*}
\section{Discussion}

Smartphone app fingerprinting is challenging because of a variety of variables that are likely to change between fingerprint building and final deployment. Such variables include device, operating system version, app version, and time. Any mismatch between variables during app fingerprinting and app identification has the potential to reduce the performance of our app classification system. To this end, we assessed how the aforementioned variables affected system performance. Apps were fingerprinted and later re-identified under a thorough suite of experimental settings.

In Table~\ref{table_baseline}, we report app classification performance when classifiers are trained and tested using datasets generated from the same dataset. In the other tests, we used completely independent datasets for training and testing. System performance when using independent sets was seen to be notably lower than the baseline experiments. This highlights the need for completely independent training and testing sets if one wants to get a more accurate estimate of the performance of an app fingerprinting system.

Training with specific app versions and device with six months between the collection of training and testing data had the highest baseline accuracy. This suggests that time (at the six month timescale) introduces the least variance in app fingerprints. This insight suggests that although the content returned by the app's servers may have changed, our models are fairly resilient to those changes and still give good performance. Our analysis on datasets collected using different devices (and operating system version) gave performance slightly lower than the previous test. This suggests that device or operating system characteristics of different devices can introduce some additional noise that affects classification performance to a small extent. Such reduction in performance is expected, since apps are known to change their behaviour depending on the version of Android operating system that they are run on. Additionally, differences in the operating system itself may also contribute additional noise that affects classifier performance.

Fingerprinting a set of apps and identifying new versions of the same apps incurred a further performance penalty. This phenomenon is not unexpected, since apps routinely receive changes to their logic during updates~\cite{Taylor:2017:UUI:3052973.3052990}, which may cause changes in their network traffic flows. However, our classification system shows that it is able to cope with such changes. This, however, motivates the need to re-fingerprint apps whenever they are updated, but suggests that old fingerprints may be useful, although presumably less so as the app receives more updates. Changing both device and app versions (and time) provided the greatest performance penalty for our classification system. This is an expected penalty since time, device, operating system version, and app versions have all changed between training and testing. Even under these most severe of constraints, our classifier was able to achieve a baseline performance~20 times that of pure random guessing.

The majority of the performance hit appears to come from so-called ambiguous flows. These flows are traffic that is similar across apps that typically comes from third-party libraries that are in common among apps. Such ambiguous traffic frustrates naive machine learning approaches, since the classifiers are given effectively the same training examples with different labels. Using a novel two-stage classification strategy with reinforcement learning, we were able to approximately double the baseline performance of our classifiers. Using the additional post-processing technique of classification validation, further accuracy could be extracted from the system, but at the expense of the number of flows that the classifiers were able to give a confident enough prediction. We remind the reader here that in app classification there is no inherent requirement to label all network flows.
\section{Conclusion}

In this paper, we extended AppScanner, a robust and scalable framework for the identification of smartphone apps from their network traffic. We thoroughly evaluated the feasibility of fingerprinting smartphone apps along several dimensions. We collected several datasets of app-generated traffic at different times (six months apart) using different devices (and Android operating systems) and different app versions. We demonstrated that the passage of time is the variable that affects app fingerprinting the least. We also showed that app fingerprints are not significantly more affected by the device that the app is installed on. Our results show that updates to apps will reduce the accuracy of fingerprints. This is unsurprising since new app versions will likely have additional features, which can affect the fingerprint recognition process. We showed that even if app fingerprints are generated on a particular device, they can be identified six months later on a different device running different versions of the same apps with a baseline accuracy that is~20 times better than random guessing. Using the techniques of ambiguity detection and classification validation, we obtained noteworthy increases in system performance. We were able to fingerprint and later re-identify apps with up to 96\% accuracy in the best case, and up to 73\% accuracy in the worst case. These results suggest that app fingerprinting and identification is indeed feasible in the real-world. App fingerprinting unlocks a variety of new challenges as it relates to user security and privacy. By continuing research in this area, we aim to better understand these challenges, so that appropriate responses can be engineered to keep users safe now and into the future.

%\section*{Acknowledgment}

\IEEEtriggeratref{29}
\bibliographystyle{IEEEtran}
\bibliography{IEEEabrv,appscanner-tifs}

\end{document}